\documentclass[12pt]{article}
\usepackage{setspace}
\usepackage[utf8]{inputenc}
\usepackage{graphicx}
\graphicspath{{Figures_preview/}}

\usepackage{amsmath}
\usepackage[normalem]{ulem}
\usepackage{cite}

\usepackage[pdf]{pstricks}
\usepackage{xr}
\usepackage[T1]{fontenc}
\usepackage{cancel}
\usepackage{makecell}
\usepackage{transparent}
\usepackage{geometry}
\usepackage{caption} 
\usepackage{lipsum}
\usepackage[many]{tcolorbox}
\usepackage{soul}
\usepackage{url}

\definecolor{darkpastelgreen}{rgb}{0.01, 0.75, 0.24}

\newcommand{\D}{\mathrm{d}}
\newtcolorbox{cross}{blank,breakable,parbox=false,
  overlay={\draw[red,line width=5pt] (interior.south west)--(interior.north east);
    \draw[red,line width=5pt] (interior.north west)--(interior.south east);}}

\title{Early stages of polycrystalline diamond deposition: Laser reflectance at substrates with growing nanodiamonds}
\author{David V\'azquez-Cort\'es, Stoffel D.\ Janssens, \\Burhannudin Sutisna, and Eliot Fried$\footnote{Corresponding author: tel: +81(098)966-1372. E-mail: eliot.fried@oist.jp}$}
\date{}

\begin{document}

\maketitle
\vspace{-1cm}
\begin{center}
Mechanics and Materials Unit (MMU), Okinawa Institute of Science and Technology Graduate University (OIST), 1919-1 Tancha, Onna-son, Kunigami-gun, Okinawa, Japan 904-0495
\end{center}
\vskip20pt


\begin{abstract}
The chemical vapor deposition of polycrystalline diamond (PCD) films is typically done on substrates seeded with diamond nanoparticles. Specular laser reflectance and a continuous film model have been used to monitor the thickness of these films during their deposition. However, most seeds are isolated during the early stages of the deposition, which questions the utility of applying such a continuous film model for monitoring deposition before film formation. In this work, we present a model based on the Rayleigh theory of scattering for laser reflectance at substrates with growing nanodiamonds to capture the early stages of PCD deposition. The reflectance behavior predicted by our model differs from that of a continuous film, which is well-described by the continuous film model. This difference enlarges as the seed density used in our model decreases. We verify this trend experimentally by depositing diamond under identical conditions on substrates with various seed densities. A relation derived from our model is used to fit reflectance data from which seed densities are obtained that are proportional to those found with electron microscopy. We also show that relying on the continuous film model for describing the early stages of deposition can result in falsely deducing the existence of incubation, and that the continuous film model can be used safely beyond the early stages of deposition. Based on these findings, we delineate a robust method for obtaining growth rates and incubation periods from reflectance measurements. This work may also advance the general understanding of nanoparticle growth and formation.
\end{abstract}
\section{Introduction}
\subsection{Motivation and goals}
According to Russel \cite{Russell1969}, the term ``incubation time'', which is used interchangeably with ``incubation period'' \cite{williams2011nanocrystalline}, ``induction time'' \cite{jiang1994nucleation}, and ``induction period'' \cite{stoner1992situ}, describes thermally activated processes that do not commence immediately after establishing a reaction temperature. A prime example of a thermally activated process is the chemical vapor deposition (CVD) of polycrystalline diamond (PCD) \cite{Butler2008}. Due to high activation energies associated with the nucleation of diamond grains on foreign substrates, relative to those of diamond deposition, incubation periods preceding deposition are observed \cite{snail1992situ,yamaguchi1994temperature, GIUSSANI2022152103}. Such incubation periods can be reduced significantly, or even eliminated entirely, by seeding diamond nanoparticles, typically detonation nanodiamonds, on substrates before deposition \cite{doi:10.1063/1.103812}. Still, the seeding step does not prevent incubation under certain deposition conditions \cite{GIUSSANI2022152103,cite-key, KROMKA20081252}. Incubation can be related to the etching of seeds \cite{williams2011nanocrystalline}, which affects the seed density.
Consequently, the morphological features of films and the physical properties such as thermal conductivity, transparency, and adhesiveness are also affected \cite{LEE20062046, ANAYA2017215, HUANG2020146733}. A fundamental understanding of incubation is therefore essential for producing PCD films with tailored properties. The occurrence of incubation is mainly deduced from monitoring deposition rates during the early stages of deposition using specular laser reflectance interferometry. This measurement method is typically based on a model that assumes continuous film interference \cite{williams2011nanocrystalline, GIUSSANI2022152103}. However, seeds are isolated during the early stages of deposition, raising questions about whether such a model can be reliably applied to processes before film formation. The goals of this work are to:
\begin{itemize}
\item Develop a semiquantitative, non-interference, and non-continuous film model for specular laser reflectance based on nanodiamond particle scattering that captures the early stages of PCD CVD on seeded substrates.
\item Test our model experimentally by depositing diamond on substrates with different seed densities and analyzing the samples with several microscopy techniques.
\item Develop a method to estimate incubation periods, which we identify with a delay in deposition and/or a reduced deposition rate during the early stages of deposition, with specular laser reflectance.
\end{itemize}
\noindent
For brevity, ``specular laser reflectance'' is shortened to ``laser reflectance''. The term ``diamond particle'' refers to a single grain or to a cluster of grains isolated from other diamond particles and the term ``seed'' refers to a diamond particle deposited on the surface of a substrate before deposition.

\subsection{Background}
The small lattice parameter and the high surface energies of diamond limit diamond heteroepitaxy to diamond-on-iridium  \cite{millan2021study,doi:10.1021/acs.cgd.9b00488,delchevalrie2021spectroscopic,cicala2013smoothness,Schreck2014, doi:10.1063/5.0024070}. An alternative method to deposit diamond on foreign substrates relies on seeding the substrate surface with diamond particles \cite{MENDESDEBARROS19961323,C.Norgard1998,AKHVLEDIANI2002545,MANDAL202179, SMITH2020620,LEE2022396}. 
Most work on seeding aims to increase the areal (seed) density with the objective of minimizing the thickness at which films become pinhole-free \cite{scorsone2009enhanced,tsigkourakos2012spin,shenderova2010seeding}. However, high seed densities are not always desirable. For example, Mandal {\it et al.}\cite{mandal2019thick} found that a low seed density reduces stress in a PCD film grown on an aluminum nitride substrate, affording the deposition of a thicker film without delamination. Tsigkourakos {\it et al.} \cite{tsigkourakos2014local} found that lowering the seed density increases the electrical conductance of boron-doped films by one order of magnitude, and Janssens {\it et al.}\cite{Janssens2022} showed, through simulations, that the seed density strongly affects the grain size distribution. These results show that the seed density is a useful parameter for tuning the properties of PCD films and is therefore explored systematically in this work. 

\begin{figure} 
\centering
\fontsize{10}{12}\selectfont
\includegraphics{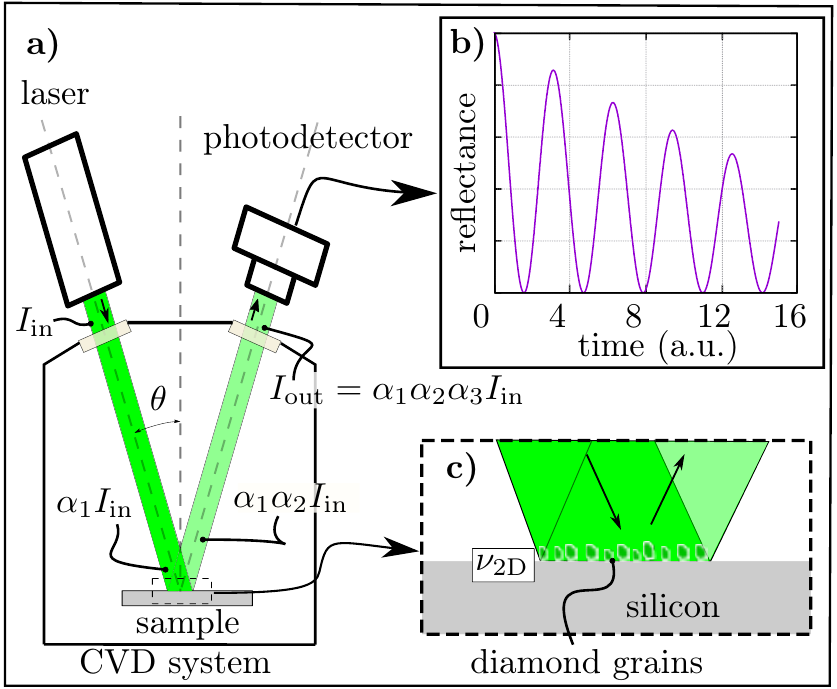}
\fontsize{12}{14.4}
\caption{ {\bf Schematic of a reflectance setup.} a) Schematic of a setup for measuring the specular laser reflectance of a sample in a chemical vapor deposition (CVD) system during deposition. A laser beam of intensity $I_\text{in}$ enters the CVD system at an angle $\theta$. After interaction with the windows and the sample, the beam exits the system with intensity $I_\text{out}$. The arrows indicate the propagation direction of the laser beam. The windows cause attenuations $\alpha_1$ and $\alpha_3$, and attenuation $\alpha_2$ is caused by the sample. The inset of a) shows a schematic of diamond particles during the outset of deposition, with seed density $\nu_\text{2D}$. b) A typical graph obtained by plotting the photodetector's signal in a) as a function of time. Regular oscillations, which are used for deposition rate estimation, are observed for thin-film interference.}
\label{fig:intro1}
\end{figure}

{\it In-situ} laser reflectance interferometry is widely used for monitoring the thickness of a nanocrystalline diamond film during deposition. This is done by shining a laser beam on a sample, collecting the specularly reflected light with a photodetector, and analyzing the output signal that the photodetector produces\cite{aida2021situ,delfaure2016monitoring,chen2020submicrometer}. Figure \ref{fig:intro1} shows a schematic of a reflectance setup installed on a CVD reactor. The film thickness and deposition rate are typically calculated from the extrema of the oscillating reflectance and the time intervals separating these extrema, respectively\cite{wu1993laser, stoner1992situ, zuiker1996situ}. The oscillations are caused by thin film interference.

In the early stage of diamond deposition on seeded silicon substrates, laser reflectance decays more slowly than expected for a continuous film\cite{zuiker1996situ,cicala2013smoothness }. Some authors have related the initial slow decay of the reflectance to a reduced deposition rate or to the varying roughness of the sample surface \cite{williams2011nanocrystalline,wu1993laser}. 
For the deposition of diamond on seeded substrates, the existence of incubation is mainly derived from the slow decay in reflectance during the early stages of deposition. However, it is not clear whether the estimation of an incubation period is strongly affected by the assumption of a continuous film. Here, we investigate this systematically by comparing laser reflectance with atomic force microscopy (AFM) and scanning electron microscopy (SEM).

Bonnot {\it et al.}\cite{bonnot1993investigation} showed that during the early stages of PCD deposition, scattering from isolated diamond seeds increases with time. This increase was measured by installing a detector away from the specularly reflected laser. The analysis of the measured scattering was done using a model based on Rayleigh scattering, assuming that scattering caused by seeds is proportional to the square of their volume. This made it possible to calculate the diamond deposition rate using scattering instead of reflectance. As demonstrated by Smolin {\it et al.}\cite{smolin1993optical}, Mie scattering, which generalizes Rayleigh scattering to large particles, can also explain the reflectance behavior during the initial stage of deposition. Smolin {\it et al.} measured reflectance by introducing a laser beam through a quartz window directed normal to the substrate surface and by measuring the specularly reflected beam intensity by a photodiode. However, the experiments reported in their work were limited by the control over seeding available at the time. Also, the implications of varying the diamond seed density on reflectance were not systematically investigated. These works demonstrate that light scattering, rather than interference, can accurately describe light--sample interactions during the early stages of PCD deposition. However, to the best of our knowledge, a relation that describes the reflectance during the early stages of PCD deposition has not yet appeared in the literature. Moreover, we are not aware of any previous attempt to include light scattering phenomena in the analysis of reflectance curves measured during incubation. We, therefore, develop an elementary model based on the Rayleigh theory of scattering to explain the behavior of laser reflectance during the early stages of PCD deposition and test the model experimentally. Due to the non-interference nature of the light--sample interaction in the early stages of PCD deposition, we avoid using the term ``laser reflectance interferometry'' and opt for the more general term ``laser reflectance''. 

\section{Model}
\label{sec:model}

\subsection {Laser beam-sample interaction}

A standard reflectance setup mounted on a CVD system is shown in Figure~\ref{fig:intro1}.  From linear optics, the attenuation $\alpha$ of a beam with intensity $I_{in}$ entering the CVD system during diamond deposition is given by
\begin{equation}
\alpha=\frac{I_\text{out}}{I_\text{in}},
\end{equation}
where $I_\text{out}$ is the intensity of the beam leaving the CVD system.
Using the same argument for the beam intensity arriving and leaving the windows and sample, the total beam attenuation can be expressed as
\begin{equation}
\alpha=\alpha_1\alpha_2\alpha_3,
\end{equation}
where $\alpha_1$ and $\alpha_3$ are the beam attenuation caused by the windows and $\alpha_2$ is the attenuation caused by the sample. The absolute specular reflectance is the ratio of the beam intensity leaving the sample to that arriving at the sample. Following this definition, $\alpha_2$ can be identified with the absolute specular reflectance of the sample consisting of diamond particles on a silicon surface. Therefore, we write $\alpha_2$ as the product of the attenuation $\alpha_{21}$ caused by scattering due to diamond particles and the absolute specular reflectance $\alpha_{22}$ of the silicon substrate: 
\begin{equation}
\alpha_2=\alpha_{21}\alpha_{22}.
\label{totalReflectance}
\end{equation}
With \eqref{totalReflectance} and $\alpha_{22}$, we then define the relative reflectance $R$ through
\begin{equation}
R = \frac{\alpha_2}{\alpha_{22}}=\alpha_{21}.
\label{normRefl}
\end{equation}

The voltage measured by the photodetector, as depicted in Figure~\ref{fig:intro1}a, is proportional to the responsivity $r$ of the detector. Dividing the voltage $V$ measured in the presence of diamond particles by the voltage $V_0$ measured for a bare silicon substrate, we obtain the relative voltage
\begin{equation}
\frac{V}{V_0}=\frac{r\alpha_1\alpha_{21}\alpha_{22}\alpha_3I_{in}}{r\alpha_1\alpha_{22}\alpha_3I_{in}} =\alpha_{21}.
\label{normalVoltage}
\end{equation}
With \eqref{normRefl} and \eqref{normalVoltage}, we find that the relative voltage $V/V_0$ equals $R$, which is entirely determined by the attenuation due to scattering by diamond particles. We also note that these results are applicable only when the specularly reflected beam intensity is much larger than the scattered intensity in the direction of the detector. This condition is fulfilled for a polished substrate with a high refractive index on which relatively small nanoparticles are present. 

\subsection{Scattering caused by a collection of particles constrained in a plane}
\label{scatt2Darray}
\begin{figure}
\centering
\fontsize{10}{12}\selectfont
\includegraphics{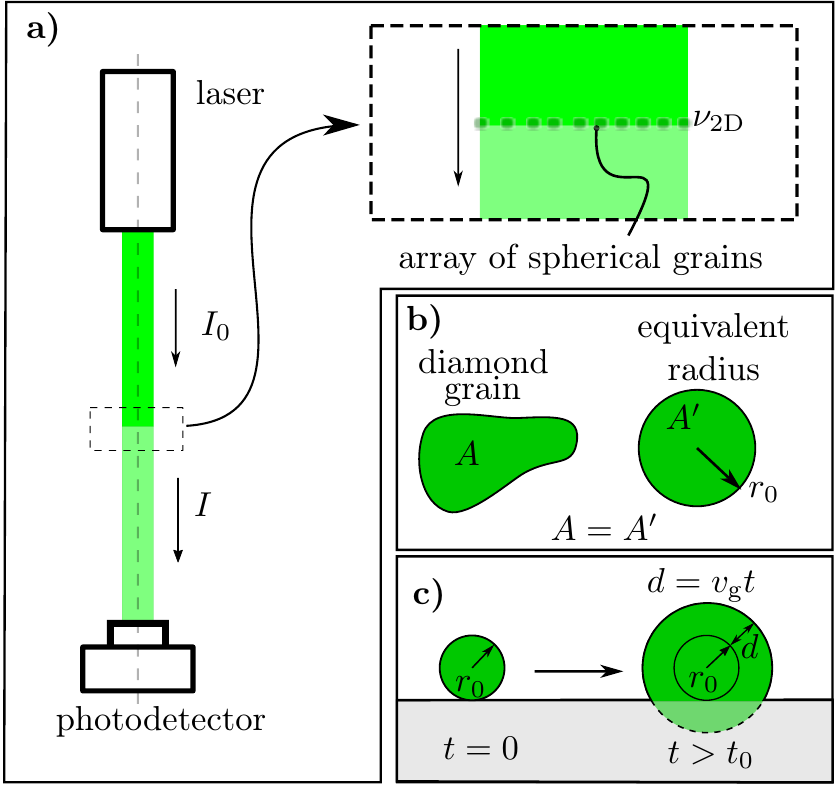}
\fontsize{12}{14.4}
\caption{{\bf Model for calculating scattering caused by growing diamond particles.} 
a) Schematic of an imaginary experimental setup to calculate the scattering caused by a collection of uniform size diamond particles constrained to a plane. The beam exits the laser with intensity $I_0$, and the intensity after interaction with the particles is reduced to $I$. The particles are assumed to be spherical and of the same radius $a$. 
b) To obtain $a$ experimentally, we measure the projected area $A$ of a diamond particle, calculate the radius $r$ of the circle with area $A'$ for which $A=A'$, and take the mean value of $r$ for all the particles. 
c) Uniform deposition on an equivalent seed with radius $a_0$ at the outset of the deposition is depicted. The radius of the particle is the sum of $a_0$ and the deposited thickness $d$. The deposited thickness is expressed as $d = v_\text{g}t$, with $v_\text{g}$ and $t$ denoting deposition rate and deposition time, respectively.} 
\label{fig:Scattering}
\end{figure}

To obtain a relation for $\alpha_{21}$, we first recognize that the intensity change, per unit length, $\D I/\D z$ of a beam with initial intensity $I_0$ traveling in a non-adsorbing host fluid containing particles with number density per unit volume $\nu$ can be expressed as
\begin{equation}
\frac{\D I}{\D z}= -I_0\nu \sigma,
\end{equation}
where $\sigma$ denotes the scattering cross-section of the particles in the fluid. The Rayleigh approximation can be written as $\lambda_0\ll2a$, where $\lambda_0$ is the wavelength of the incoming beam in the host fluid and $a$ the radius of the suspended particles. In the Rayleigh approximation, the scattering cross-section $\sigma_\text{R}$ is given by 
\begin{equation}
\sigma_\text{R}=\frac{8\pi}{3}\Big(\frac{2\pi n_g}{\lambda_0}\Big)^4a^6\Big(\frac{m^2-1}{m^2+2}\Big)^2,
\label{RayCross}
\end{equation}
with $m=n_p/n_g$, where $n_p$ and $n_g$ are the refractive indices of the particles and the host fluid respectively \cite{cox2002experiment}. When the particles are constrained to a plane, as shown in Figure~\ref{fig:Scattering}, the change $\Delta I$ in intensity $I$ for a beam traveling through the particles is written as
\begin{equation}
\Delta I=-I_0\nu_{\text{2D}}\sigma_\text{R},
\label{eq:deltaI}
\end{equation}
where $\nu_{2D}$ is the density of particles. The beam intensity $I$ after interaction with the particles is  thus given by
%
%
\begin{equation}
I=I_0+\Delta I =I_0(1-\nu_{\mskip1mu 2D}\sigma_\text{R}).
\label{scatt}
\end{equation}
A beam with oblique incidence at angle $\theta$ interacts with an effective density of particles given by $\nu_\text{2D}/\cos\theta$. Replacing $\nu_\text{2D}$ in \eqref{scatt} with $\nu_\text{2D}/\cos\theta$ and dividing by $I_0$, we observe that the attenuation of this beam after interaction with a collection of particles constrained on a plane can be expressed as
\begin{equation}
\frac{I}{I_0}=1-\beta \frac{\nu_{\text{2D}}}{\cos\theta}a^6,
\label{I}
\end{equation}
where $\beta$ is defined by
\begin{align*}
\beta=\frac{8\pi}{3}\Big(\frac{2\pi n_\text{g}}{\lambda_0} \Big)^4\Big(\frac{m^2-1}{m^2+2}\Big)^2.
\end{align*}

Relation \eqref{I} is valid for a collection of isolated, relatively small, and uniformly sized spherical particles. However, in processes of interest here, seeds are randomly positioned on the silicon substrate after seeding, and their size is not uniform. Also, diamond particles grow homoepitaxially from the seeds and coalesce during deposition, which affects $\nu_\text{2D}$. Without performing a complete statistical analysis of the growth and coalescence of diamond particles, we model the relative reflectance $R$ in the initial stages of diamond deposition using $\alpha_{21}=I/I_0$ and assuming that:
\begin{enumerate}
\item The diamond particles are sufficiently small to ensure that the Rayleigh approximation applies.
\item Particle clustering, which reduces particle density, can be ignored.
\item Diamond particles can be represented by uniformly sized spheres of radius $a$. Obtaining $a$ experimentally is described in Figure~\ref{fig:Scattering}b and its caption. 
\item The particle size increases uniformly. Consequently, the radius $a$ increases as 
\begin{equation}
a = a_0 + d, 
\end{equation}
with $a_0$ and $d$ denoting the mean equivalent radius of seeds and the deposited thickness, respectively, as illustrated in Figure~\ref{fig:Scattering}c. The thickness $d$ is expressed as 
\begin{equation}
d= v_\text{g}t, 
\label{depThick}
\end{equation}
with $v_\text{g}$ and $t$ denoting the deposition rate and deposition time, respectively. 
\end{enumerate}

With these provisions, the length $a$ in \eqref{I} represents the mean equivalent radius of a collection of diamond particles. Using \eqref{I} in \eqref{normRefl}, we obtain
\begin{equation}
R=1-\beta\frac{\nu_{\text{2D}}}{\cos\theta}a^6.
\label{reflectanceRad}
\end{equation}
With the fourth assumption, the particle radius at time $t$ is given by 
\begin{align}
a &=a_0+v_\text{g}t.
\label{eq:radiusGrowth}
\end{align}
Substituting \eqref{eq:radiusGrowth} in \eqref{reflectanceRad}, we obtain the representation for $R$, namely
\begin{equation}
R=1-\beta\frac{\nu_{\text{2D}}}{\cos\theta}(a_0+v_\text{g}t)^6,
\label{eq:reflectanceTime}
\end{equation}
which facilitates comparisons with data from reflectance measurements. 
%

\section{Experiment}

Diamond deposition was performed on $2\times2~\text{cm}$ silicon substrates cleaved from 100~mm diameter phosphorus-doped silicon wafers with a resistivity of $1\text{--}10~\Omega \text{cm}$ obtained from \emph{Electronics and Materials Corporation, Limited}. After cleaving, the substrates were cleaned by a sequential ultrasonic bath in deionized (DI) water, acetone, isopropanol, and DI water for 5~min each.  The substrates were then immersed for 10~min in Semicoclean solution obtained from \emph{Furuuchi Chemical Corporation},  rinsed in DI water, and dried with high purity nitrogen.
After the cleaning process, the substrates were exposed to an air plasma in a PR200 Plasma reactor obtained from \emph{Yamato Scientific Corporation, Limited} to improve the wettability of their surfaces. The plasma power and exposure time were set to 170~W and 60~s, respectively, and the airflow rate was kept at 100~sccm. Immediately after the plasma treatment, the substrates were transferred to a spin coater, where they were seeded with suspensions of various nanodiamond concentrations. Cleaning and seeding were done without pause to avoid airborne surface contamination affecting the seeding \cite{Pobedinskas2021}. The seeding procedure was done by drop casting $150~\mu\text{L}$ of nanodiamond suspension on an immobile substrate. 1~min after drop casting the suspension, the sample was spun at a rotation speed of 6000~rpm for 30~s while flushing the wetted surface with DI water for the first 5~s of spinning. 

Nanodiamond suspensions were made by dispersion of NanoAmando Hard Hydrogel detonated nanodiamond powder obtained from \emph{NanoCarbon Research Institute Corporation, Limited} in 0.2~L of DI water at a proportion of 0.5~g/L, and ultrasonication of the mixture with an Epishear probe sonicator with tip diameter 3.2~mm and length 4.5~cm. This was achieved using an ultrasonic transducer with a power of 100~W and a frequency of 20~kHz. To avoid overheating the system, an on/off cycle of 1~s/1~s was fixed for 90~min. The obtained suspension was centrifuged at 15000~rpm for 30~min to remove large agglomerates of more than 140~nm. The supernatant was recovered and diluted with DI water to obtain concentrations of 10\%--60\% with respect to the as-centrifuged supernatant concentration of 100\%.

Diamond was deposited on the seeded substrates using an SDS6500X microwave plasma-assisted CVD system from \emph{Cornes Technologies.} For all samples, PCD was deposited with a plasma power of 1500~W, a hydrogen flow rate of 288~sccm, a methane flow rate of 12~sccm, and a chamber pressure of 15~torr. This was done using programmed recipes to ensure repeatability. The microwaves and the plasma in the CVD chamber heated the substrate. The temperature of each sample was found to stabilize at $530\pm8^\circ$C, as measured by a TMCX-HLN infrared thermometer from \emph{Japan Sensor Corporation} with emissivity set to $0.6$. The deposition was monitored \emph{in-situ} by a homemade laser reflectance setup consisting of a 1~mW power semiconductor laser emitting at 532 nm that was directed to the sample through a viewport in the CVD chamber at angle $\theta$ equal to approximately 20$^{\circ}$. A schematic of this setup is provided in Figure~\ref{fig:intro1}a. A silicon photodetector collected the reflected light from the substrate surface. A laser line filter centered at 532~nm with an FWHM of 1~nm was positioned between the viewport and the silicon photodetector to avoid plasma emission reaching the photodetector. 

The measured deposition rate was $0.6\pm0.05~\text{nm/min}$ obtained by growing a reference sample for 5~h. The thickness of the calibration layer was 180~nm, as measured with an optical nanoguage C13027-11 obtained from \emph{Hamamatsu Photonics} and corroborated by cross-section scanning electron microscopy (SEM) observation. Seeded substrates were characterized using a JEOL JSM-7900F scanning electron microscope obtained from \emph{JEOL}. Analysis of the SEM images was automated by a Python (\url{https://www.python.org}) script that makes use of the Image-J (\url{https://imagej.net/Fiji}) bandpass filter to denoise the original images and the OpenCV ``findContours()'' function for the particle analysis. The number density $\nu_{\text{2D}}$ is the number of isolated particles recognized by an in-house Python script and averaged over five images at 50 000X magnification from different portions on the substrates. After the deposition, the samples were characterized by atomic force microscopy (AFM) on an Agilent 5500 scanning probe microscope obtained from \emph{Keysight Technologies} using silicon probes with 7~nm nominal radius, 14~$\mu$m nominal cantilever length, a nominal cantilever resonance frequency of 300~KHz and a 26~N/m spring constant. AFM image analysis and profile extraction were done using Gwyddion (\url{http://gwyddion.net}). Statistical analysis of particle size was done by particle marking using the ``watershed'' method \cite{klapetek2003atomic}.  

\section{Results and discussion}
\subsection{Predictions of the model}
\begin{figure}
\centering
\fontsize{10}{12}\selectfont
\includegraphics{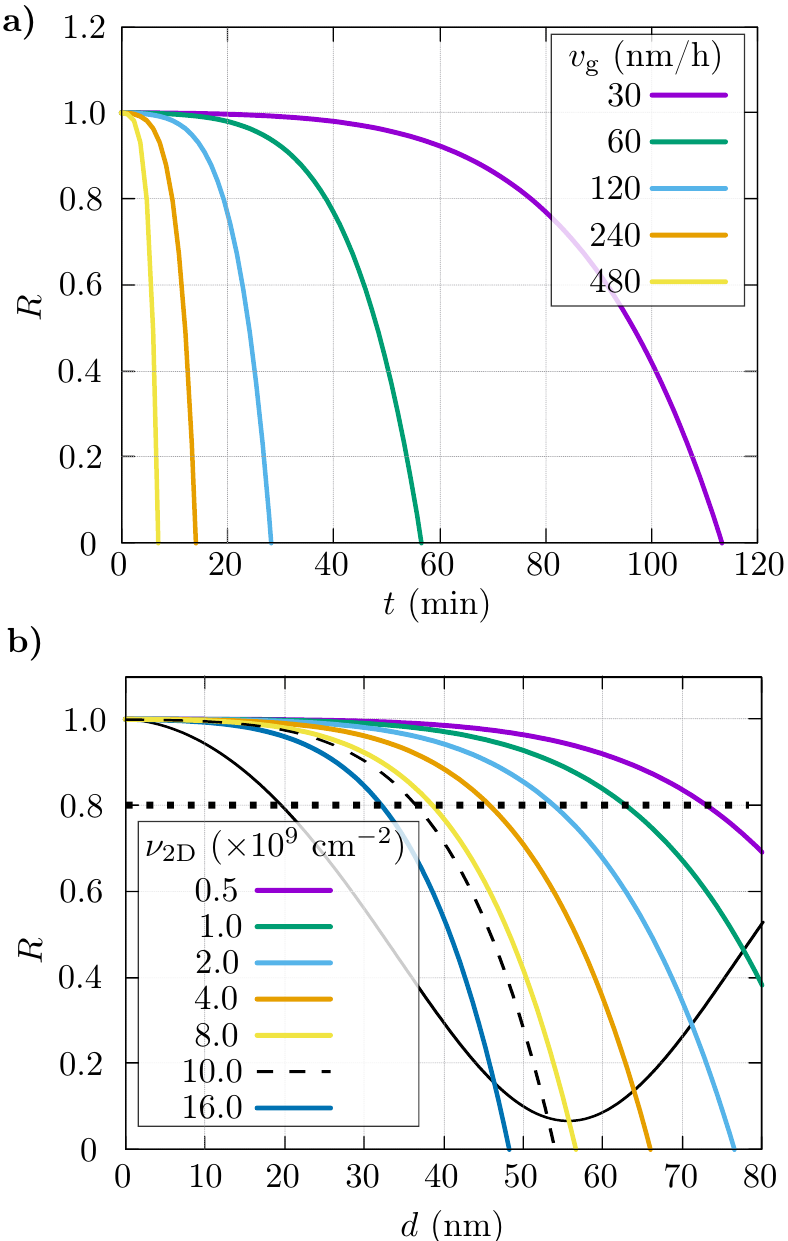}
\fontsize{12}{14.4}
\caption{{\bf Simulated reflectance during diamond deposition.} a) Relative reflectance $R$ calculated using \eqref{eq:reflectanceTime} for diamond deposition on seeded substrates as a function of deposition time $t$ for various deposition rates. b) $R$ calculated using \eqref{eq:reflectanceThick} as a function of deposited thickness $d$ for various values of $\nu_{\text{2D}}$. Radius $a_0$ was set to 20~nm for all calculations. The black curve in b) corresponds to the normalized reflectance for a continuous diamond film calculated by the transfer-matrix method \cite{orfanidis2002electromagnetic}.} 
\label{fig:AnalysisEquation}
\end{figure}

To understand the implications of our model, we plot the relative reflectance $R$ versus $t$ using the Rayleigh scattering relation \eqref{eq:reflectanceTime} for various choices of $v_\text{g}$ with $a_0$ = 20 nm and $\nu_\text{2D} = 1\times 10^{10}\ \text{cm}^{-2}$. 
Figure~\ref{fig:AnalysisEquation}a shows that $R$ decays more slowly as $v_g$ decreases. To avoid confusing this effect with the presence of incubation, we use \eqref{depThick} and \eqref{eq:reflectanceTime} to express $R$ in terms of $d$:
\begin{equation}
R=1-\beta\frac{\nu_{\text{2D}}}{\cos\theta}(a_0+d)^6.
\label{eq:reflectanceThick}
\end{equation}
Using \eqref{eq:reflectanceThick}, the reflectance data in Figure~\ref{fig:AnalysisEquation}a collapses on the dashed curve in Figure~\ref{fig:AnalysisEquation}b, where $R$ is plotted as a function of $d$. In Figure~\ref{fig:AnalysisEquation}b $R$ is also plotted for other values of $\nu_{\text{2D}}$. The black curve corresponds to $R$ of a continuous diamond film calculated by the transfer-matrix method \cite{orfanidis2002electromagnetic}. The colored curves exhibit less decay than the black curve as $d$ increases. Guided by the horizontal dotted line at a fixed value of $R$, and assuming our model correctly predicts $R$ as a function of $d$, we deduce that an experimenter would underestimate the value of $d$ using the continuous film model during the early stages of deposition. Consequently, film thickness and deposition rate are also underestimated with the continuous film model, and the presence of an incubation period might falsely be deduced. Low values of $\nu_\text{2D}$ cause less scattering than high values, which explains the slower decay of $R$ as $\nu_\text{2D}$ decreases.  

In a review on nanocrystalline diamond by Williams \cite{williams2011nanocrystalline}, $R$ is plotted as a function of $t$ for diamond deposition on seeded substrates. The reflectance curves exhibit a decay that reduces as the methane concentration in the gas phase reduces. This trend is reasoned to be caused by an extension of the incubation period as the methane concentration is reduced. However, by assuming that $v_\text{g}$ is proportional to the methane concentration, a possible alternative explanation is that a reduced methane concentration produces a lower deposition rate that delays the reflectance decay. The reality is probably a combination of both descriptions since a plasma with lower methane concentration might etch grains{, leading to} incubation. To resolve the ambiguity regarding the existence of incubation, we refer to a method delineated in our previous work \cite{GIUSSANI2022152103} for estimating incubation periods. In Section~\ref{beyond}, we extend this method.

\subsection{Experimental verification of the model}
\label{growthFitEtc}
We test our model by comparing the relative reflectance $R$ during diamond deposition on seeded substrates with various values of $\nu_{\text{2D}}$. The values of $\nu_{\text{2D}}$ and $a_0$ were obtained by SEM analysis.
The Rayleigh scattering relation \eqref{eq:reflectanceTime} was fitted to the reflectance curves obtained during diamond deposition using the parameters $\nu_{\text{2D}}$, $a_0$, and $V_0$. The values of $\nu_{\text{2D}}$ and $a_0$ were compared with those obtained from SEM. After deposition, the surface morphology of the samples was analyzed by AFM. Finally, the height distributions of diamond particles were correlated to the reflectance measurements.

\subsubsection{Seeding} 
\label{seeding} 
\begin{figure}
\centering
\fontsize{10}{12}\selectfont
\includegraphics{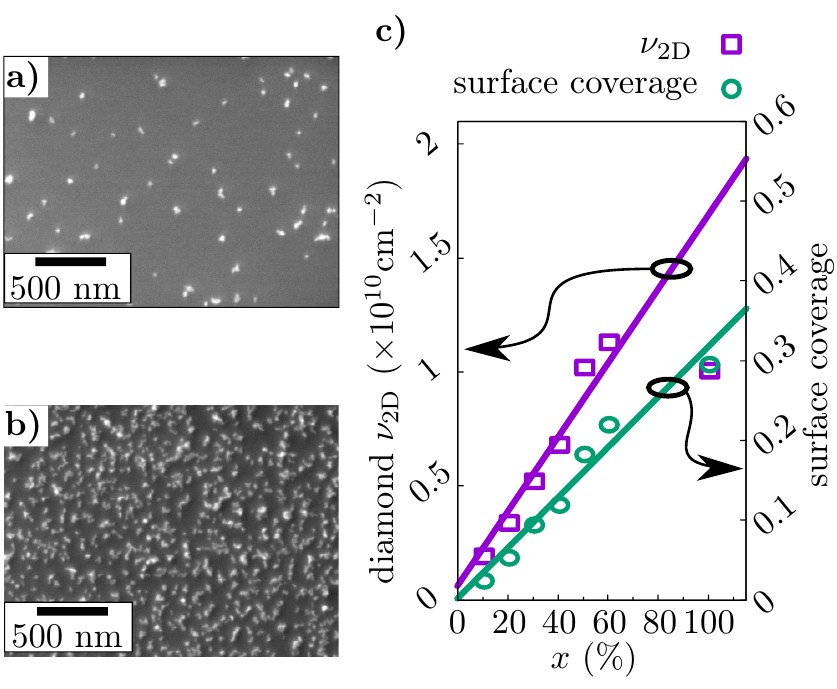}
\fontsize{12}{14.4}
\caption{{\bf Seed density.} a) SEM image of a sample seeded using a suspension of concentration $x =10$\%. b) SEM image of a sample seeded with an $x = 100$\% suspension. c) Seed density $\nu_{\text{2D}}$ and surface coverage as function of $x$. Density $\nu_{\text{2D}}$ is defined as the number of isolated diamond particles in the image, divided by the total area. For the linear fit the value of $\nu_\text{2D}$ for $x$ $= 100$\% is not considered.}
\label{fig:SEM}
\end{figure}

The SEM images in Figures~\ref{fig:SEM}a--b show substrates seeded using suspensions of concentration $x = 10$\% and $x = 100$\%, respectively, from which we observe that the surface coverage and $\nu_\text{2D}$ increase as $x$ increases. Figure~\ref{fig:SEM}c reveals that this increase is linear for the surface coverage in the full range, and linear for $\nu_\text{2D}$ up to 60\%. 
We speculate that the deviation of $\nu_\text{2D}$ from the linear behavior at 100\% is caused by agglomeration and the limited resolution of the SEM, which make it challenging to resolve seeds in close proximity. 
These results show that our seeding procedure enables control over $\nu_\text{2D}$. A complete set of images of seeded substrates is shown in Figure~S1 in the supporting information. The figure also includes images after particle analysis with contours drawn around the detected seeds. 

In the work of Tsigkourakos~\emph{et al.} \cite{tsigkourakos2014local} the surface coverage was found to vary non-linearly for suspension concentrations in the range of 0.002~g/l to 2~g/l. Recognizing that nanodiamonds in suspension tend to aggregate relatively fast at low concentrations \cite{Mchedlov_Petrossyan_2015}, we hypothesize that nanodiamond agglomerates were seeded in that work. Scorsone \emph{et al.}~\cite{scorsone2009enhanced} obtained good linearity between $\nu_\text{2D}$ and nanodiamond concentration in the suspension. Still, the suspension preparation required the addition of polyvinyl alcohol and ultrasonication at $40 ^{\circ}$C for 2~h. In the work described here, we simply diluted a stock suspension immediately before seeding to avoid dilution-induced agglomeration. 
 %
 
\subsubsection{Reflectance during the early stages of diamond deposition}
\label{reflectanceGrowth}

\begin{figure}
\centering
\fontsize{10}{12}\selectfont
\includegraphics{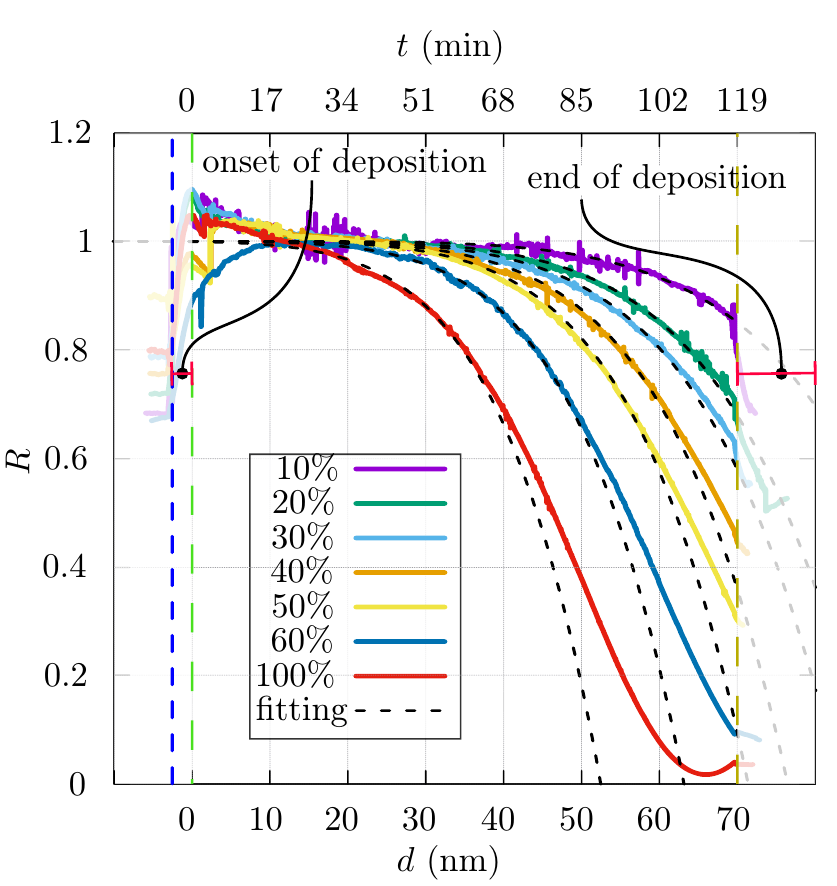}
\fontsize{12}{14.4}
\caption{{\bf Experimental and simulated reflectance curves.} Laser reflectance $R$ as a function of the deposition time $t$ and the deposited thickness $d$. $R$ is monitored during the deposition of diamond on silicon substrates seeded with nanodiamond suspension of concentrations ranging from 10\% to 100\% 
relative to the stock suspension. The dashed black curves correspond to fitting \eqref{eq:reflectanceTime} to the experimental curves. The green and dark yellow vertical dashed lines indicate the start and end of the deposition, respectively. The blue dashed line indicates the time at which the plasma is ignited.} 
\label{fig:reflectanceAll}
\end{figure}

The relative reflectance $R$ measured during diamond deposition on samples seeded with suspensions of various $x$ is shown in Figure~\ref{fig:reflectanceAll} as a function of $d$. The dashed black lines in Figure~\ref{fig:reflectanceAll} are obtained by fitting the alternative representation \eqref{eq:reflectanceTime} for $R$ to the data, and are discussed hereinafter. Three time intervals can be identified in each plot: (1) the onset of deposition, (2) the deposition, and (3) the end of the deposition. The onset of deposition comprises the interval starting at the ignition of the plasma and finishing when the substrate temperature reaches $530^{\circ}$C. This time interval, typically about 5~min, begins and ends in the positions indicated by blue and green vertical dashed lines, respectively. The end of deposition starts when the plasma power is deactivated. During this time interval, the substrate experiences a sudden drop in temperature, which is driven by the cooled stage of the CVD system. The deposition interval occurs between the onset of the deposition interval and the end of the deposition interval. A relatively small $v_\text{g}$ of 0.6~nm/min was selected to guarantee that deposition outside the deposition interval can be safely neglected. 
 
As predicted by our model, the decay of $R$ as a function of $d$ is inversely proportional to $\nu_{\text{2D}}$. In contrast to findings reported in the existing literature, our measurements show that a diminished decay of $R$ is not per se related to incubation. Evidence for incubation is mainly based on literature \cite{williams2011nanocrystalline,stoner1992situ} that describes polycrystalline diamond deposition by the bias-enhanced nucleation method. 
Stoner~\cite{stoner1992situ} found that in the initial stage of diamond deposition, by this method, incubation is observed, during which no diamond deposition occurs. This has been corroborated by Jiang \emph{et al.}~\cite{jiang1994nucleation} using AFM measurements which showed that diamond nuclei were not created until after 6.5~min of plasma exposure. In the works of Stoner and Jiang \emph{et al.}, incubation is ascribed to diamond nucleation that starts after the silicon substrate is saturated with carbon. For polycrystalline diamond deposition on seeded substrates under harsh conditions, we previously showed that incubation is present due to seed etching~\cite{GIUSSANI2022152103}. However, under mild conditions similar to those used in this work, incubation has never been corroborated, to the best of our knowledge~\cite{ARNAULT20081143,STANISHEVSKY20151403}.

Quantitative results are obtained by non-linear least-square fitting of \eqref{eq:reflectanceTime} with the values $v_\text{g}=0.6$~nm/min, $n_{\text{p}}=2.41$, $n_{\text{g}}=1$, and $\lambda_0=532$~nm and fitting parameters $\nu_\text{2D}$, $a_0$, and $V_0$. The fitting ranges are manually adjusted and are listed in Table \ref{table}.
From the dashed curves in Figure~\ref{fig:reflectanceAll}, the fits for samples seeded using low-concentration diamond suspensions are good over the entire deposition interval. However, the fits and the experimental data diverge for high concentrations before the deposition ends. This can be attributed to the proximity of neighboring particles for samples with high seed densities. When the radius of a diamond particle exceeds half of the distance to its nearest neighbor, they coalesce and form a larger particle. This phenomenon has two consequences: 
\begin{enumerate}
\item The diamond particle density $\nu_{\text{2D}}$ will be reduced due to coalescence.
\item Coalesced particles may be larger than the wavelength for which the Rayleigh scattering approximation holds, leading to significant discrepancies between the model and the experimental data. 
\end{enumerate}
Once all diamond particles coalesce to form a continuous layer of polycrystalline diamond, which we called the closed-film threshold in our previous work \cite{Janssens2022}, the reflectance signal can be described using the transfer matrix method \cite{orfanidis2002electromagnetic}. By visual inspection, it is also evident that the fitting curves coincide with experimental data up to a reflectance value of $0.7\pm0.1$, which we interpret as the empirical reflectance limit above which our model can be applied with confidence.
\begin{table}
\fontsize{10}{12}\selectfont
\centering 
\begin{tabular}{c c c c c c}
\makecell{sample\\ \%}	&\makecell{$\nu_{\text{2D}}$ (SEM)\\ $\times$ $10^{9}$ cm$^{-2}$} & \makecell{$a_0$ (SEM)\\ nm} &\makecell{$\nu_{\text{2D}}$ (fitting)\\ $\times$ $10^{9}$ cm$^{-2}$} &\makecell{$a_0$ (fitting)\\ nm} 	&\makecell{Fitting range\\ $d$ nm}\\
\hline
\hline

10		&1.9				&19			&0.9		&10	&10--69\\
20		&3.4				&20			&1.6		&13	&10--67\\
30		&5.2				&22			&1.8		&14	&10--65\\
40		&6.8				&21			&2.8		&14	&10--63\\
50		&10.2			&21			&3.9		&14	&10--61\\
60		&11.3			&22			&5.5		&18	&10--41\\
100		&10.1			&27			&8.4		&23	&10--29\\
\end{tabular}
\caption{Nanodiamond seed densities and mean equivalent radius, measured by SEM analysis and estimated by fitting \eqref{eq:radiusGrowth} to the experimental data in Figure \ref{fig:reflectanceAll}. The fitting range for all samples is also included.}
\label{table}
\fontsize{12}{14.4}
\end{table}
\begin{figure}
\centering
\fontsize{10}{12}\selectfont
\includegraphics{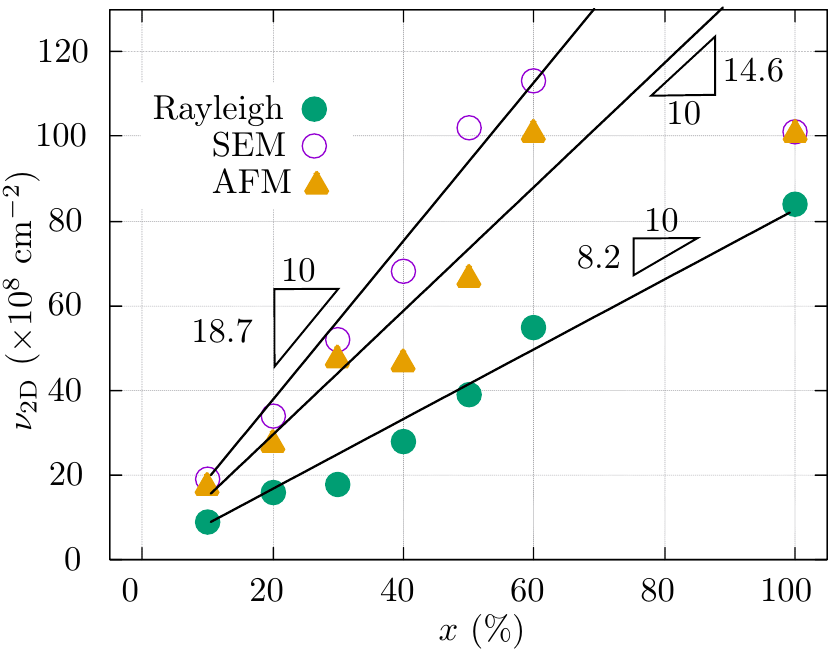}
\fontsize{12}{14.4}
\caption{{\bf Estimated diamond particle densities.} Diamond particle density $\nu_{\text{2D}}$, as a function of $x$, obtained by (a) fitting the Rayleigh scattering relation \eqref{eq:reflectanceTime} to the reflectance curves displayed in Figure~\ref{fig:reflectanceAll}, (b) the SEM image analysis that is delineated in Section~\ref{seeding} (before deposition), and (c) AFM particle marking that is delineated in Section~\ref{sec:AFM} (after deposition).  For the linear fit, the values of $\nu_\text{2D}$ at 100\% obtained by SEM and AFM are not considered.} 
\label{fig:SEMvsFIT}
\end{figure}
Values for $\nu_{\text{2D}}$ obtained by fitting the curves in Figure~\ref{fig:reflectanceAll} are shown in Table~\ref{table} and values for $\nu_{\text{2D}}$ estimated by SEM analysis in Section~\ref{seeding} are also shown for comparison. 
Figure~\ref{fig:SEMvsFIT} shows that except for the sample grown on the substrate seeded with the 100\% concentration suspension, the values for $\nu_{\text{2D}}$ obtained by SEM analysis are proportional to those obtained by fitting using the Rayleigh scattering relation \eqref{eq:reflectanceTime} with a proportionality constant of 0.44. 

\subsubsection{Seed density versus growth rate}
\label{sec:AFM}
Our model hinges on the assumption that the deposition rate $v_\text{g}$ is independent of the particle density $\nu_{\text{2D}}$. Particle height statistics were performed on AFM data to assess the validity of this assumption. We obtained statistical information from the height of the particles by the ``watershed'' method \cite{klapetek2003atomic}. This method allows for marking particles based on height local minima/maxima in an AFM image; thereby, coalesced particles forming clusters or forming a film can be identified as individual particles. 
\begin{figure}
\centering
\fontsize{10}{12}\selectfont
\includegraphics{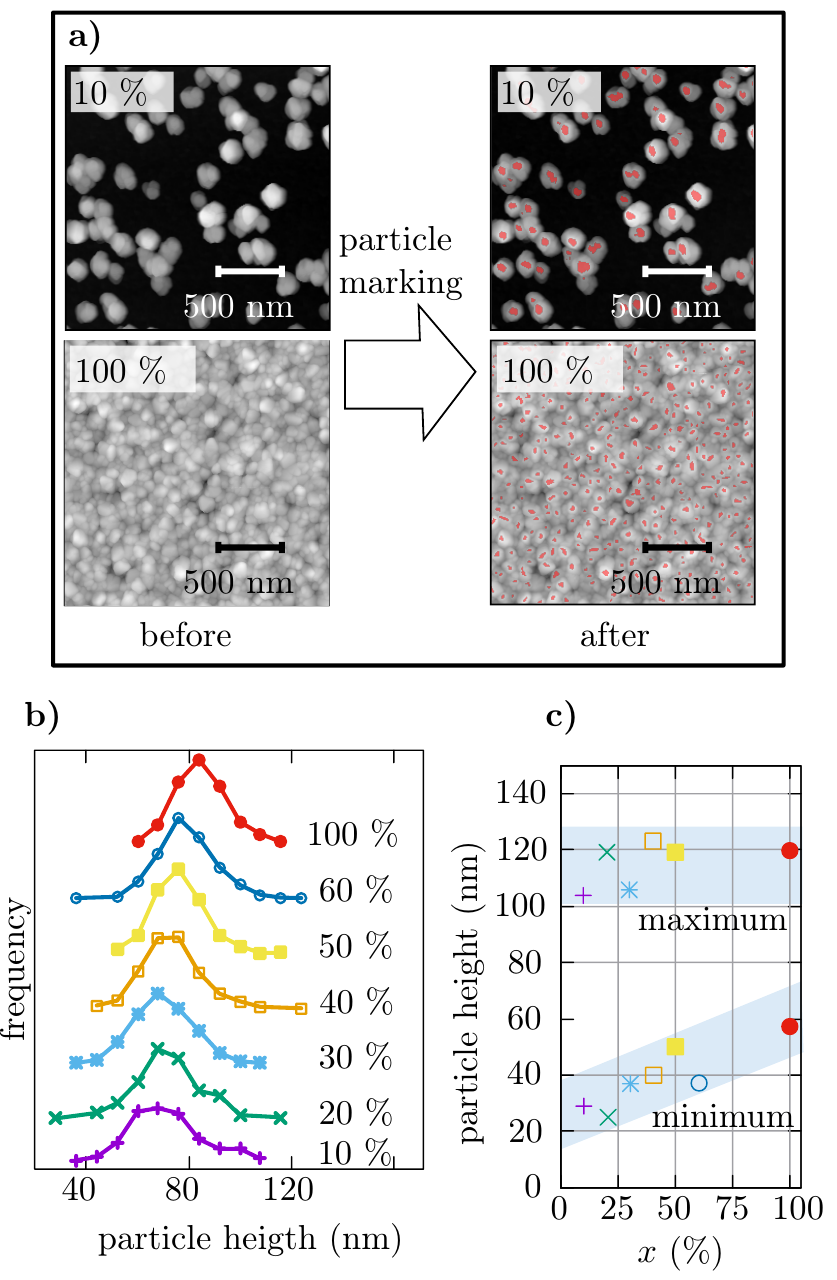}
\caption{{\bf AFM particle analysis.} a) AFM images of samples seeded with suspensions of concentration $x=10\%$ and $x=100\%$, before and after applying the watershed method for marking particles. b) Height distribution and c) the maximum and the minimum height of particles found in the analysis of two AFM images of each sample.}
\label{fig:AFMgrain}
\fontsize{12}{14.4}
\end{figure}
The particle density, the height distribution, the maximum particle height, and the minimum particle height for each sample were obtained by analyzing AFM data from two areas measured at different positions. In Figure~\ref{fig:SEMvsFIT}, the values of $\nu_{\text{2D}}$ obtained by SEM and AFM, before and after deposition, respectively, are plotted as functions of the concentration $x$ of diamond particles in solution. The smaller slope obtained from AFM is caused by grain burying, as explained in the following paragraph.

Figure~\ref{fig:AFMgrain}a contains examples of AFM images before and after particle marking. A complete set of images can be found in Figure~S2 of the supporting information. 
The particle height distributions are shown in Figure~\ref{fig:AFMgrain}b, and the maximum and minimum height as a function of $x$ are provided in Figure \ref{fig:AFMgrain}c. From this data, we find that (i) the particle height at the maximum frequency increases as $x$ increases, (ii) the maximum height ($114\pm10$~nm) is not significantly dependent on $x$, and (iii) the minimum height increases as $x$ increases.  These findings can be explained as follows. During deposition, smaller particles are buried by larger particles\cite{Drift2014APG}. Burying occurs more frequently as $\nu_\text{2D}$ increases because the mean distance between particles decreases as $\nu_\text{2D}$ increases. Consequently, the particle height at the maximum frequency and the minimum particle height shift towards larger values. In this scenario, the particle of maximum height is not affected by burying and its height is independent of $\nu_{\text{2D}}$ provided that $v_\text{g}$ is constant for all the samples, as we assume to be so. 


\subsection{Seed density versus mean film thickness}
\label{beyond}
Since surface roughness decreases after film closure \cite{wu1993laser} and the maximum grain height is independent of $\nu_{\text{2D}}$, as shown in Section~\ref{sec:AFM}, the mean film thickness should become less dependent on $\nu_{\text{2D}}$ as the deposited thickness increases. As a consequence, we hypothesize that the $R$ curves for samples grown on a substrate with dissimilar $\nu_{\text{2D}}$ should show extrema at values of $d$ that become more similar as $d$ increases. 
\begin{figure}
\centering
\fontsize{10}{12}\selectfont
\includegraphics{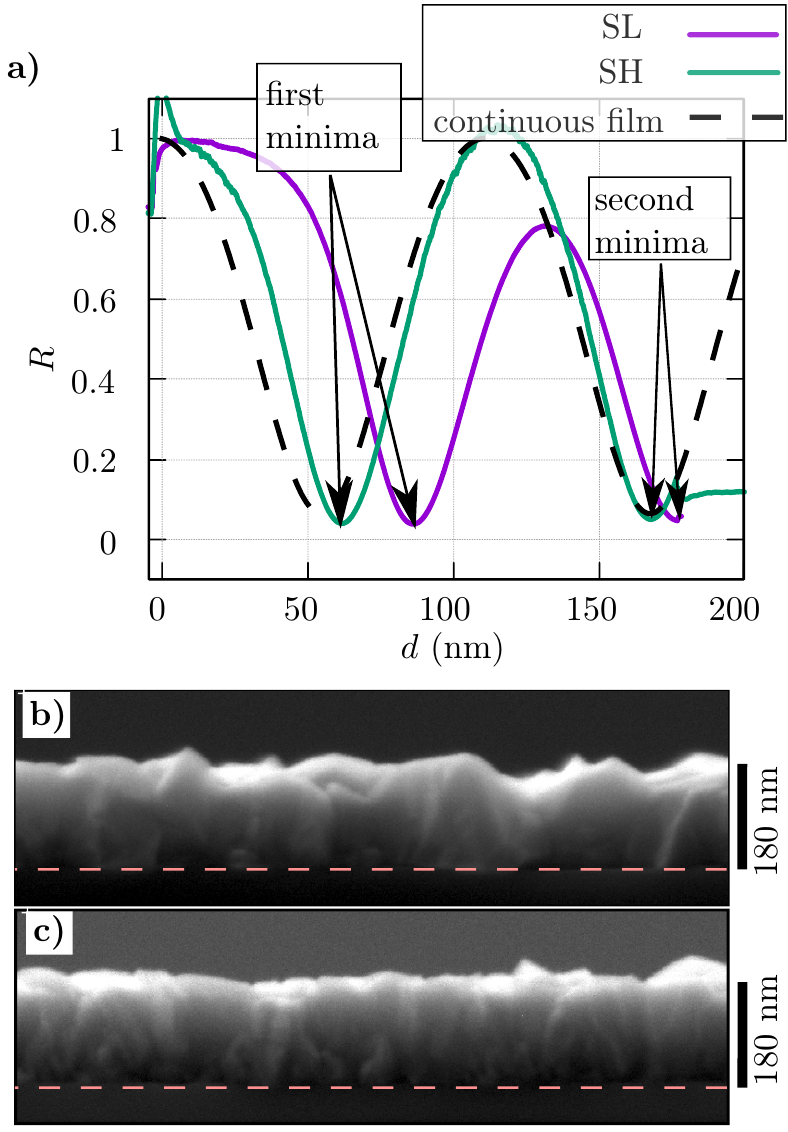}
\caption{{\bf Film growth after the early stages of deposition.} a) Reflectance data measured during diamond deposition plotted versus deposited thickness $d$ of samples SL and SH with seed density $\nu_{\text{2D}}$ equal to $1.00\times10^{10}~\text{cm}^{-2}$ and $2.95\times10^{10}~\text{cm}^{-2}$, respectively. The plots show that after depositing 180~nm of diamond, the positions of the minima tend to the same value of $d$.  b) and c) show SEM cross-sections of SL and SH, respectively. Both images show that SL and SH have similar thicknesses (180~nm), but the RMS surface roughness of SL appears larger than that of SH. A red dashed line in b) or c) indicates the substrate--diamond interface. 
SH shows a similar thickness of 180~nm. Higher RMS surface roughness in sample SL than in sample SH is also observed. The red line in the image indicates the substrate--diamond interface.}
\label{fig:AddExp}
\fontsize{12}{14.4}
\end{figure}
Using Figure~\ref{fig:AddExp}a, we test this hypothesis by comparing $R$ as a function of $d$ for samples SL and SH with $\nu_{\text{2D}}=1.00\times10^{10}~\text{cm}^{-2}$ and $\nu_{\text{2D}}=2.95\times10^{10}~\text{cm}^{-2}$, respectively, as measured by SEM. While the $R$ curve for SL exhibits the first minimum at approximately $d=86.0$~nm, the first minimum for the $R$ curve of SH is at approximately $d=61.5$~nm. On the other hand, the position of the second minimum is about $d=178$~nm and $d=168$~nm for SL and SH, respectively. The difference in $d$ for the second minimum is only 10~nm, which is considerably smaller than the 24.5~nm observed for the first minimum and, therefore, supports our hypothesis.  

Figures \ref{fig:AddExp}b and \ref{fig:AddExp}c show the SEM cross-section of SL and SH, respectively. From these cross-sections, we deduce that the mean film thickness is similar in both samples, which is in line with SL thickness 182~nm and SH thickness 184~nm that we obtained by an optical nanoguage. From these cross sections, we also observe that the surface of SL is rougher than that of SH. We verify and quantify this by analyzing AFM data, obtained from two $5\times5~\mu\text{m}^2$ areas measured at different positions, for each sample. AFM images from these samples and their profiles along the selected lines are shown in Figure~S3 of the supporting information. The result of the analysis is 14.6~nm and 11.0~nm RMS surface roughness for SL and SH, respectively. This difference in surface roughness might contribute to the difference in $d$ for the second minima in Figure~\ref{fig:AddExp}a. In addition to surface roughness, we hypothesize that voids at the substrate-film interface might also contribute to this difference. 

The results presented in the present section also show that even for a relatively low value of $\nu_\text{2D}$, the continuous film model can provide a reasonable approximation of the mean film thickness beyond the early stages of deposition.

\subsection{Estimation of incubation periods by laser reflectance}

\begin{figure}
\centering
\fontsize{10}{12}\selectfont
\includegraphics{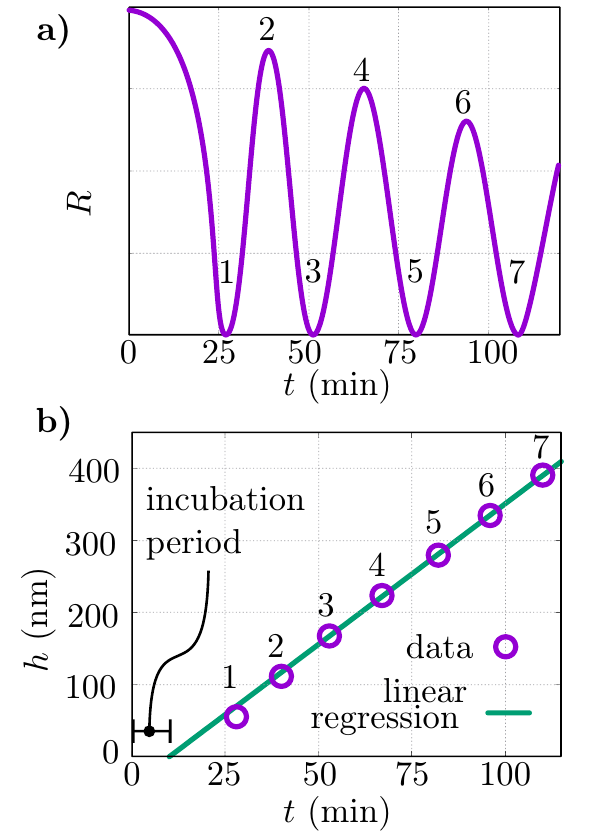}
\caption{{\bf Schematic reflectance to depict the method to estimate the incubation period.} a) Typical reflectance $R$ curve obtained during diamond deposition on seeded substrates as a function of time $t$ where extrema are numbered. b) Incremental thickness $h$ as a function of $t$ for the curve in a). The values of $h$ are calculated with \eqref{incrementalThickness}. }
\label{refOnlyIncubation}
\fontsize{12}{14.4}
\end{figure}

To systematically obtain an incubation period and the deposition rate by laser reflectance, we use the findings presented in this work to extend a previously proposed method~\cite{GIUSSANI2022152103}. To quantify the incubation period, we collect the values of $t$ where the extrema in $R$ occur. A schematic of such a curve is provided in Figure~\ref{refOnlyIncubation}a. Subsequently, we estimate the incremental thickness $h=l\Delta h$, which we define as the thickness at the $l^\text{th}$ extremum, calculated for a continuous film. The thickness $\Delta h$ is a constant that can be calculated with the relations
\begin{equation}
\Delta h = \frac{\lambda_0}{4n_\text{Dia}\cos\gamma},  
\qquad 
\gamma = \arcsin\left(\frac{\sin\theta}{n_\text{Dia}}\right),
\label{incrementalThickness}
\end{equation}
where $n_\text{Dia}$ is the refractive index of diamond, assuming that the surrounding medium is vacuum with a refractive index equal to unity. 
Corresponding pairs of $h$ and $t$ are then plotted as illustrated in Figure~\ref{refOnlyIncubation}b. Finally, we perform the linear regression with all the data points except those corresponding to the first two extrema. From Section \ref{beyond}, we derive that the values of $h$ corresponding to the first two extrema are not well-described by the continuous film model. These extrema should therefore be avoided, especially for low values of $\nu_\text{2D}$. The intercept with the $t$-axis and the slope, both obtained from the linear regression, are the incubation period and growth rate, respectively. 

In the event that few or no extrema are observed, more sophisticated characterization techniques such as ``specific zone marking''~\cite{delfaure2016monitoring}, that enable the observation and measurement of localized diamond particles before and after diamond deposition, can be used.

\section{Conclusions}
Based on Rayleigh scattering caused by growing diamond particles, we presented a simple model to describe specular laser reflectance in the early stages of polycrystalline diamond deposition. The reflectance behavior predicted by our model differed from that of a continuous film, which is well-described by a continuous film model, and this difference enlarged as the seed density used in our model decreased. These predictions were tested and confirmed by \emph{in-situ} specular laser reflectance measurements of diamond deposition on silicon substrates with various seed densities. A relation for reflectance found in our model was fitted to the experimental reflectance data, and the seed density obtained by fitting was found to be proportional to the seed density obtained by scanning electron microscopy analysis. Post-deposition atomic force microscopy data analysis showed that the deposition rate is independent of the seed density, which supports an assumption made in our model. Other deposition experiments showed that the continuous film model can be used safely beyond the early stages of deposition. We also showed that relying on the continuous film model for describing the early stages of deposition can result in questionable conclusions regarding the presence of incubation or the duration of incubation periods. Inspired by this, we proposed a method for estimating incubation periods together with a deposition rate, based on laser reflectance measurements. 
\begin{table}[h!]
\caption{Nomenclature }
\begin{tabular}{l p{13cm} }
Symbol	&Description	\\
\hline \hline
$a$	&Mean equivalent radius\\
$a_0$	&Mean equivalent radius of diamond seeds\\
$\alpha$	&Total beam attenuation caused by the CVD system	\\
$\alpha_1$	&Beam attenuation caused by entering the CVD system\\
$\alpha_2$	&Beam attenuation caused by the sample\\
$\alpha_{21}$	&Beam attenuation caused by the diamond particles on the silicon substrate\\
$\alpha_{22}$	&Beam attenuation caused by the total reflectance of the silicon substrate\\
$\alpha_3$	&Beam attenuation caused by exiting the CVD system\\
$\beta$	&$\displaystyle\frac{8\pi}{3}\Big( \frac{2\pi n_\text{g}}{\lambda_0} \Big)^4\Big( \frac{m^2-1}{m^2+2} \Big)^2$\\
$d$	&Deposited thickness\\
$\sigma$	&Scattering cross section\\
$\sigma_\text{R}$	&Rayleigh scattering cross section\\
$I_0$	&Laser beam intensity before interaction with small particles\\
$I$	&Laser beam intensity\\
$I_\text{in}$	&Beam intensity entering the CVD system\\
$I_\text{out}$	&Beam intensity leaving the CVD system\\
$l$	&Number of extrema\\
$\lambda_0$	&Beam wavelength in the medium\\
$m$	&Ratio $n_\text{p}/n_\text{g}$\\
$n_\text{g}$	&Refractive index of the medium\\
$n_\text{p}$	&Refractive index of the particles\\
$\nu$	&Number density per unit volume\\
$\nu_\text{2D}$	&Nanodiamond seed density\\
$R$	&Relative reflectance\\
$r$	&Equivalent radius\\
$t$	&Time\\
$\theta$	&Angle of incidence\\
$V$	&Voltage measured during diamond growth at time $t$\\
$V_0$	&Voltage for a laser beam reflected on a bare silicon substrate\\
$v_\text{g}$	&Deposition rate\\
$x$	&Suspension concentration\\
\end{tabular}
\fontsize{12}{14.4}
\end{table}
%

\section{Acknowledgements}
The authors thank the Okinawa Institute of Science and Technology Graduate University (OIST) for the financial support, with subsidy funding from the Cabinet Office, Government of Japan. The authors also thank T.~Sasaki, from the Scientific Imaging Section at OIST, for the assistance and technical guidance on SEM measurements. S.~D.~J.\ and E.~F.\ acknowledge JSPS KAKENHI funding Grant Number JP21K03782, and B.~S.\ acknowledge JSPS KAKENHI (Grants-in-Aid for Early-Career Scientists) Grant Number JP20K15347.

\bibliographystyle{unsrt}
\bibliography{bibliography}
\end{document}